\documentclass[journal,twoside,web]{ieeecolor}
\usepackage{tmi}
\usepackage{cite}
\usepackage{amsmath,amssymb,amsfonts}
\usepackage{algorithm}
\usepackage{algorithmic}
\usepackage{graphicx}
\usepackage{textcomp}

\def\BibTeX{{\rm B\kern-.05em{\sc i\kern-.025em b}\kern-.08em
    T\kern-.1667em\lower.7ex\hbox{E}\kern-.125emX}}
\begin{document}
\title{SOUP-GAN: Super-Resolution MRI Using Generative Adversarial Networks}
\author{Kuan Zhang, Haoji Hu, Kenneth Philbrick, Gian Marco Conte, Joseph D. Sobek, Pouria Rouzrokh, Bradley J. Erickson
\thanks{Corresponding author: Bradley J. Erickson (e-mail: bje@mayo.edu).}
\thanks{K. Zhang, G. M. Conte, J. D. Sobek, P. Rouzrokh and B. J. Erickson are with the Department of Radiology, Mayo Clinic, Rochester, MN. }
\thanks{H. Hu is with the Department of Computer Science \& Engineering, the University of Minnesota, Minneapolis, MN.}
\thanks{K. Philbrick is with Google LLC, Mountain View, CA. Work was conducted while employed at Mayo Clinic.}}

\maketitle

\begin{abstract}
There is a growing demand for high-resolution (HR) medical images in both the clinical and research applications. Image quality is inevitably traded off with the acquisition time for better patient comfort, lower examination costs, dose, and fewer motion-induced artifacts. For many image-based tasks, increasing the apparent resolution in the perpendicular plane to produce multi-planar reformats or 3D images is commonly used. Single image super-resolution (SR) is a promising technique to provide HR images based on unsupervised learning to increase resolution of a 2D image, but there are few reports on 3D SR. Further, perceptual loss is proposed in the literature to better capture the textual details and edges than using pixel-wise loss functions, by comparing the semantic distances in the high-dimensional feature space of a pre-trained 2D network (e.g., VGG). However, it is not clear how one should generalize it to 3D medical images, and the attendant implications are still unclear. In this paper, we propose a framework called SOUP-GAN: $\bold{S}$uper-resolution $\bold{O}$ptimized $\bold{U}$sing $\bold{P}$erceptual-tuned $\bold{G}$enerative $\bold{A}$dversarial $\bold{N}$etwork (GAN), in order to produce thinner slice (e.g., high resolution in the `Z' plane) medical images with anti-aliasing and deblurring. The proposed method outperforms other conventional resolution-enhancement methods and previous SR work on medical images upon both qualitative and quantitative comparisons. Specifically, we examine the model in terms of its generalization for various SR ratios and imaging modalities. By addressing those limitations, our model shows promise as a novel 3D SR interpolation technique, providing potential applications in both clinical and research settings.  
\end{abstract}


\section{Introduction}
\label{sec:introduction}

Medical imaging modalities (e.g., MRI, CT and ultrasound imaging) are widely used for clinical diagnosis and research \cite{Essential_Physics}. They reveal internal anatomy and can provide quantitative measurements of the human body. Acquisition of a medical image depends on both the technical quality of the instrument and the conditions of the scan (e.g. patient properties, nature of the image being acquired), and in most cases, involves compromise, i.e., choosing a shorter acquisition time in MRI, reducing the radiation dose in CT, and using lower ultrasound power levels in ultrasound imaging. Thus, the acquired images may have limited spatial resolution, most commonly lower in the ‘z-plane’. For instance, MRI using 2D spin-echo technique can provide good in-plane resolution, but typically have a high thickness for a clinically reasonable acquisition time, constraining the possibility of capturing important signals of lesions between the slices. Moreover, the thick-slice images produce poor quality oblique or perpendicular reconstructions. There is active demand in practice to generate HR images in order to relieve the burden of patients, to reduce examination costs and to decrease motion-induced artifacts.	  

One of the main approaches to accelerate MRI acquisition is through under-sampling the data in k-space, which inevitably results in aliasing artifacts by violating the Nyquist-Shannon sampling criteria \cite{Nyquist}, and then performing reconstruction of the under-sampled MRI, by solving an ill-posed inverse problem from an optimization point of view \cite{Regularization_methods}. Currently, parallel imaging (PI)  and compressed sensing (CS) \cite{CS} are widely used in clinical practice. While the former method adopts multiple independent channels to simultaneously receive each view of the tissue area and utilizes software tools such as SENSE \cite{SENSE} or GRAPPA \cite{GRAPPA} to combine the multiple signals, the latter provides a framework for imaging reconstruction by adding a regularization term subject to prior sparsity constraints into the optimization problem. Beyond the conventional techniques, deep learning methods for MRI reconstruction have been extensively studied to learn the relationship between high-quality ground truths as target and the corresponding under-sampled images as input \cite{DAGAN, SANTIS, MoDL, Cyclic_Loss}. More advanced networks and algorithms are proposed following the idea of incorporating more ``physics" from conventional methods (CS and PI), e.g., the development of variational networks \cite{Variational, endtoend, Kevin_dual-space}.     

Single-image SR focuses on the 2D spatial domain, and refers to the process of generating HR images from low-resolution (LR) images. Deep learning based SR models have been actively explored and achieve state-of-the-art performance on natural image datasets \cite{SRCNN, ESPCN,SRGAN, EDSR, ESRGAN}. One important breakthrough is the perceptual loss function \cite{PerLoss, PerLoss_1, PIRM}, which is proposed to tackle the issue of blurring generated by optimization of pixel-wise loss functions such as MSE. Together with GAN loss, perceptual loss has the potential to create images that are visually indistinguishable from the ground truth images. 

Basic SR techniques have been introduced to MRI. Chaudhari et al. \cite{SR_MRI}  implements convolutional neural networks (CNNs) to learn residual information between HR thin-slice images and LR thick-slice images using MSE loss. A self-supervised anti-aliasing algorithm  is developed by \cite{SMORE, Zhao} to resolve the LR out-of-plane slices by learning from the HR in-plane slices, thus requiring no external training dataset. High-quality results are reported by the SR MRI approaches in literature \cite{SR_mri_1, SR_mri_2,SR_mri_3, SR_mri_4, SR_mri_5}. However, there are certain limitations that preclude their application in real-world scenarios. Given the blurred textures trained by MSE loss, perceptual loss with GAN seems necessary in the single-image SR model, but it is unclear how to generalize the pre-trained 2D model based perceptual loss to 3D medical images. In modern medical image deep learning tasks, data augmentation is often applied during data preprocessing when the acquired data is insufficient. The adaptation of the SR MRI model to address a range of SR scenarios such as various sampling ratios, acquisition processes, or medical modalities is still limited. As a result, interpolation is still used instead of SR for routine data augmentation.

In this study, we proposed a novel algorithm termed SOUP-GAN: $\bold{S}$uper-resolution $\bold{O}$ptimized $\bold{U}$sing a $\bold{P}$erceptual-tuned $\bold{GAN}$ network for generating HR thin-slice images. Our main contributions are:
\begin{itemize}
\setlength{\itemsep}{0pt}
\item We generalize the application of perceptual loss previously defined on 2D pre-trained VGG onto 3D medical images. Together with GAN, the new scheme with 3D perceptual loss significantly improves the perceived image quality over the MSE trained results. 
\item We develop a multi-scale SR architecture that works for various sampling factors with an overall satisfactory performance. 
\item Two criteria are proposed accounting for different acquisition protocols. Without examining the acquisition process carefully, the extracted datasets will not provide accurate mapping from LR slices into HR slices, and therefore may not work well for all the SR tasks.
\item We evaluate the feasibility of designing one model that works for all the medical images. The proposed model is applied to datasets from other imaging modalities, e.g., T2-weighted MRI and CT. Combining the above aspects, our model shows potential to be applied for the novel SR interpolations on medical images. 
\end{itemize}

\section{Method}
\label{sec:2}
\subsection{Residual CNNs for 3D SR-MRI}

Generally, image reconstruction is the transformation of signals acquired to images that are meaningful to humans and can be formulated as the multi-dimensional linear system:
\begin{equation}
E x = y + \varepsilon,
\label{eq:1}
\end{equation}
which tries to generate the desired images $x = {x_i, i\leq N}$ based on the observed signals $ y= {y_i, i\leq n}$ with $N$ and $n$ as the numbers of desired and observed slices (images in the Z-dimension), respectively. $E$ is the encoder operator, representing various image generation or transformation tasks, e.g., an identity operator with and without convolutions respectively for image deblurring and denoising, image-space uniform under-sampling for SR, local masking operators for inpainting and k-space under-samples for MRI reconstruction. $\varepsilon$ is the coherent noise by the measurement. 

SR tasks on 3D MRI specifically can be viewed as an under-sampling encoder trained to generate HR slices from LR slices. While Eq.~\ref{eq:1} itself is ill-posed, it can be turned into an optimization problem:
\begin{equation}
\hat{x} = \arg\min \left(\frac{1}{2} ||Ex-y||^2_2 \right),
\label{eq:2}
\end{equation}
and solved by supervised learning. Given a set of HR ground-truth examples x, the corresponding LR images y can be obtained through the encoder operator $E = (x\ast \kappa)\downarrow_s$ that typically applies the anti-aliasing filtering with a subsequent down-sampling by the sampling factor $s$ on the ground truth slices. The training dataset is composed of the ground-truths $x$ and their associated under-sampled $y$, on which a generator network $G_\theta$ is proposed and trained to approximate the function $F$, the inverse form of the encoder $E$, such that $x= F(y)$.

The designed generator $G_\theta$ consists of a pre-upsampling layer by cubic interpolation together with a residual network of 3D convolutional layers, or 3D U-net, on which the data patches can be efficiently trained. For the task involving a large training dataset, the so-called RRDB (Residual in Residual Dense Block) introduced in ESRGAN \cite{ESRGAN} is adopted to better resolve the residual spatial details, and to alleviate the burden of occupied memory through such a post-upsampling scheme.

\begin{figure*}
\center{\includegraphics[width=0.9\textwidth]{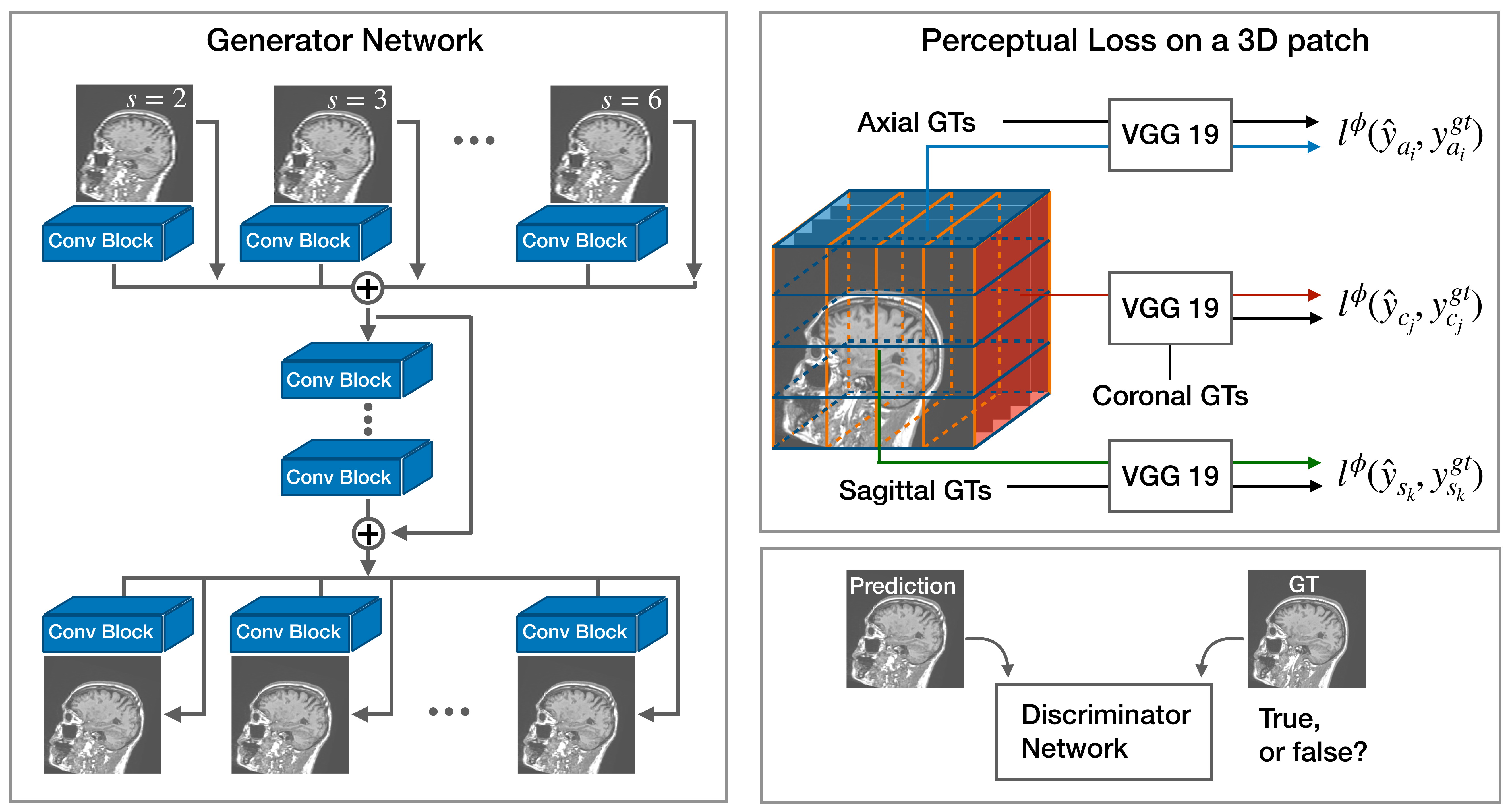}}
\caption{Architecture of our multi-scale SR network with GAN. 3D perceptual loss is calculated by inputting all the axial, sagittal and coronal surfaces of the 3D data into the pre-trained VGG, and comparing the differences between predictions and ground-truths in the high-dimensional feature space. 
}
\label{fig:1}
\end{figure*}

As an outline of the proposed framework SOUP-GAN, Algorithm \ref{alg1} is provided with pseudocode to summarize the training process including multiple processing steps, which are depicted by Fig.~\ref{fig:1} and explained in the later subsections.   

\begin{algorithm} 
	\caption{Training the proposed framework SOUP-GAN.} 
	\label{alg1} 
	\begin{algorithmic}
		\REQUIRE HR 3D MRI volume. 
		\ENSURE 1) \textbf{Data preprocessing:}
		\STATE Create LR volume by data-preprocessing following either the \textit{thin-to-thick} or \textit{thin-to-thin} criteria. 		
		\ENSURE 2) \textbf{Prepare training dataset:}
		\STATE Partition the LR data as input and HR data as ground truth into pairs of $32\times32\times32$ patches.
		\ENSURE 3) \textbf{Multi-scale SR network:}		
		\STATE Input LR patches to the multi-scale SR network with an appropriate module entrance and exit according to the associated sampling factor $s$.
		\ENSURE 4) \textbf{3D perceptual loss with GAN:}
		\STATE Based on the pre-trained MSE results, further tune the model by employing the 3D perceptual loss with GAN.
		
	\end{algorithmic} 
\end{algorithm}

\subsection{Data Preprocessing}

Data acquired for this study are 3D MRI datasets. Nine (9) T1-weighted brain MRI volumes with 1 mm isotropic pixel spacing in both the in-plane and out-of-plane directions were cropped and stacked into one $256\times256\times1170$ volume. Those thin slices were HR ground-truth images. The LR volume ($256\times256\times234$) was created by the data-preprocessing methods explained later. Both the input and ground-truth datasets were then partitioned into pairs of patches. While larger volumetric patches potentially capture better semantic features \cite{receptive_field}, we selected a patch size of $32\times32\times32$ to reduce computational demands. All patch pairs were shuffled, and distributed into training, validation and test datasets at a ratio of $8 : 1 : 1$.   

The specific way to generate the input datasets really depends on the acquisition protocols and the up-sampling factor $s$ selected for the SR task.  We define two different criteria accounting for the situations broadly categorized in the acquisition process, because without taking care of these variations, the extracted datasets will not provide accurate mapping from LR slices into HR slices, and therefore will fail to work well for various resolution-enhanced tasks in practice. In detail, the acquisition process is categorized into two types: 1) \textit{thin-to-thick}; and 2) \textit{thin-to-thin}. 

The first type assumes the same FOV of the 3D data (i.e., same Z coverage), and is subject to the selection of the slice thickness, which determines how much tissue is to be covered within each slice. Thicker slices can achieve more efficient scans without slice gaps or spatial aliasing, but may cause image blurring as partial volume effects. Correspondingly, the thick-slice representation of the HR ground-truth data is obtained by the finite impulse response (FIR), low-pass filtering with the filter coefficients as $b = 1/s$ to mimic the averaging and anti-aliasing characteristic with the high thickness, followed by a subsequent down-sampling with the factor $s$. 

The second type is to expedite the scans by discretizing FOV in Z (i.e., same 2D X-Y resolution). This is useful for the acquisition protocols focused on the in-plane resolution without considering the perpendicular (Z) planes. Thin slices with large separation are acquired efficiently. As a result, they are insufficient to cover the whole FOV in Z direction, therefore inevitably introducing spatial aliasing in the perpendicular planes \cite{Aliasing}. One clinical example is for diagnosing interstitial lung disease (ILD) in which HR CT data ($\sim$ 1mm thin slice) is acquired at 10 mm intervals, since the disease is fairly diffuse. Correspondingly to this type of clinical acquisition as well as imaging interpolation in deep learning tasks, the LR representation of the HR ground-truth data is obtained directly by the down-sampling with the factor s. For other cases between the first and second types of data preprocessing discussed here, a Gaussian filter with a tuned value of the standard deviation can be adopted. Finally, the input dataset to the network is obtained by a cubic spline interpolation with s on the LR representations that are obtained from the algorithms based on the thin-to-thick or thin-to-thin criteria.

\subsection{Multi-scale Model for SR Interpolation}

\begin{figure}
\center{\includegraphics[width=0.4\textwidth]{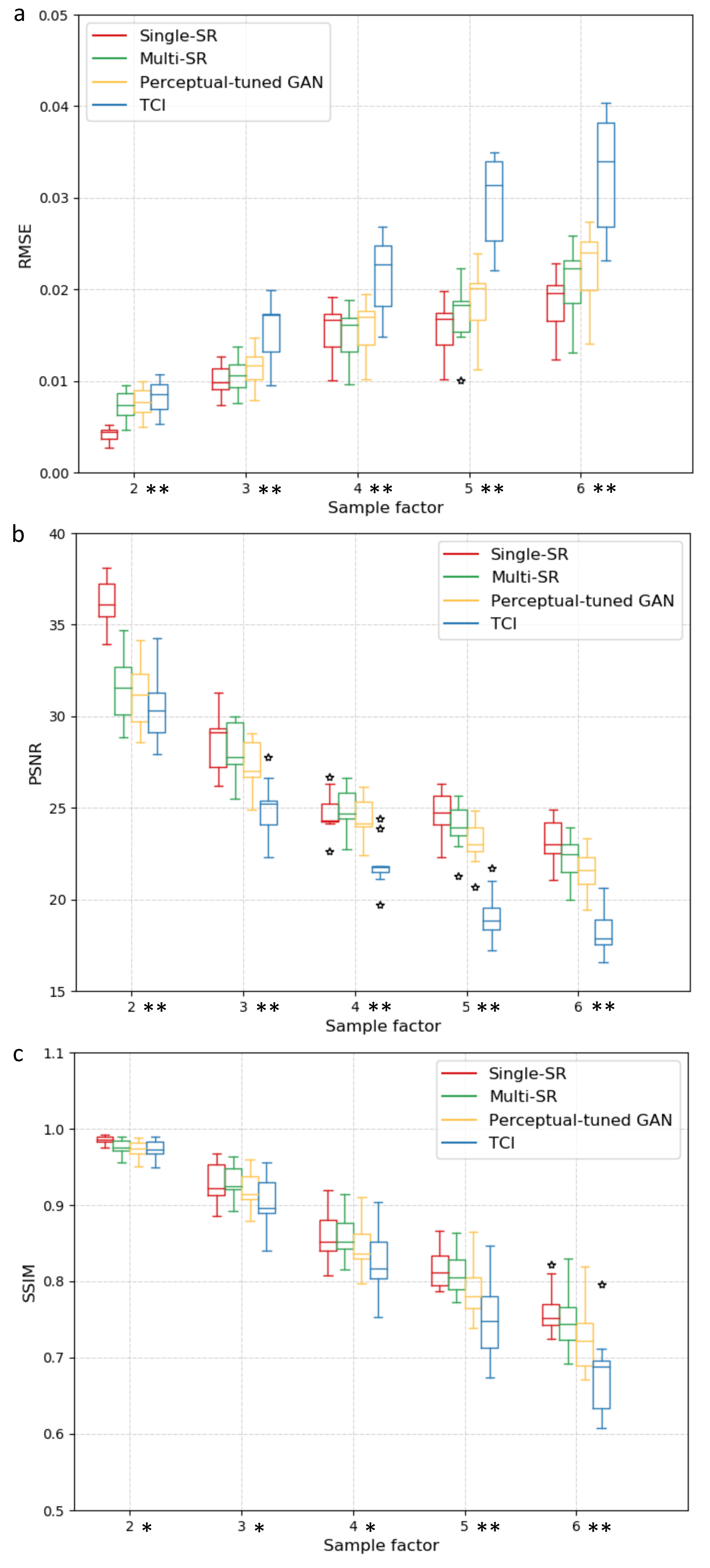}}
\caption{Quantitative comparisons of measures: (a) RMSE, (b) PSNR and (c) SSIM in terms of the sampling factor s between the single-scale model, the multi-scale model, perceptual-tuned GAN (SOUP) and cubic interpolation. Statistical significance is indicated on the x-axes with * for $p < 0.05$ and ** for $p <0.001$.
}
\label{fig:2}
\end{figure}

Supervised medical deep-learning tasks (e.g., classification, detection and segmentation) are based on the acquired images and associated labels. The accuracy of the model inference is not only decided by the design of the networks, but more importantly relies on the data size and quality. Data augmentation is often performed when the training data is insufficient, imbalanced or incompatible in shape. Conventional interpolations (such as linear, square, and cubic) are widely used and can provide approximate intermediate slices. However, when the sampling factor $s$ increases, the interpolated new images become more blurred and further away from the optimal resolution. Whether or not healthier and more realistic images could be generated seems to become the bottle-neck to further improve the model performance through data augmentation. From the observation on SR tasks, we design a multi-scale model to provide image interpolation with superior resolution, upon a user-selected sampling factor $s$.  

 In our multi-scale architecture, the single-scale model described earlier provides the starting point for learning the residual between LR and HR slices. Pre- and post- processing modules are added respectively at the top and the bottom layers, to deal with the differential information between scales. As a result, we construct one model that is able to provide predictions for multiple cases with different sampling factors $s \in \{2,3,4,5,6\}$, see Fig.~\ref{fig:2}, while most of the model parameters are shared by the backbone across the scales.       

For fractional values of the sampling factor $s$ between the selected integers, we incorporate them by interpolating the corresponding parameters of the two neighboring networks to derive an interpolated model:
\begin{equation}
\theta_G^s = (1-\alpha) \theta_G^m + \alpha \theta_G^{m+1},
\label{eq:3}
\end{equation}
where $s$ is between the integer $m$ and $m+1$ with the ratio $\alpha = s - m$. In conclusion, the proposed multi-scale SR model together with the parameter interpolation scheme provides an efficient tool to perform the image interpolation with higher resolutions than conventional interpolation methods.  

\subsection{3D Perceptual Loss}

In previous single image SR tasks studied on natural images, it was observed that training based on a pixel-wise loss, e.g., L1, MSE, or PSNR can achieve relatively high metric scores, but often results in overly-smooth SR images with blurring, since the perceptual quality of textures, edges and high-frequency details are ignored by the pixel-wise losses. Based on the idea of being closer to semantic similarity, Gatys et al. \cite{PerLoss} proposes perceptual loss, which is defined by evaluating the Euclidean distances between the feature maps extracted from a pre-trained VGG network instead of low-level pixel-wise error. For instance, perceptual loss based on the high-level representations at the $l$-th layer of the VGG19 network \cite{VGG} is formulated as:
\begin{equation}
L_{per} = \frac{1}{hwc} \sum_{i,j,k} \left( \phi^l_{i,j,k}(x) - \phi^l_{i,j,k}(F(y))\right)^2,
\label{eq:4}
\end{equation}
where $h$, $w$ and $c$ are the height, width and channel number of the representation on the layer $l$, respectively.

It is worth noting that SR MRI tasks deal with 3D medical images, while the VGG network takes 2D images as input. To generalize the single image SR with perceptual loss into 3D, we propose to adopt the original slices together with adjacent planes of the ground truth volume $x$ and the predicted volume $F(y)$. That is, all the axial, sagittal and coronal images of the 3D data will be imported into the pre-trained VGG and contribute to the perceptual loss calculation (see Fig.~\ref{fig:1}). In this way, the 3D MRI volume is perceptually resolved in terms of projected slices along the three principal directions. We also note that we used the deep intermediate layer VGG19-54 before activation rather than after activation. 

\begin{figure*}
\center{\includegraphics[width=0.8\textwidth]{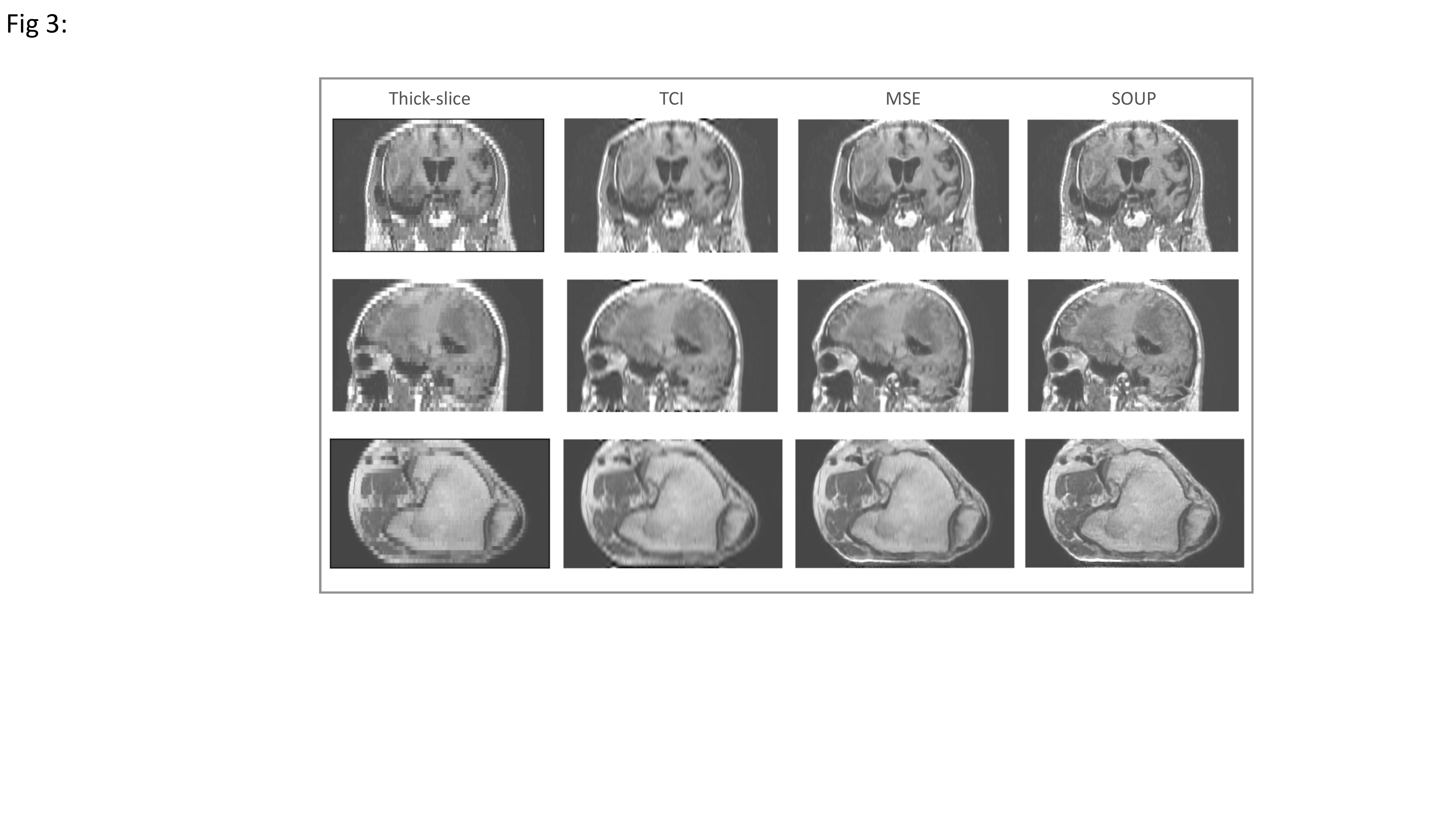}}
\caption{Test examples on brain and knee MRI showing perpendicular reconstructions from original thick slices, using tricubic interpolation (TCI), MSE and GAN using perceptual loss (SOUP). We note \cite{SR_MRI} proposed a single-scale residual-based 3D convolutional neural network with MSE, similar to the third column here. Our SOUP approach better resolves the perceptual details and is more generally applicable to various sampling factors and different imaging modalities.
}
\label{fig:3}
\end{figure*}

\begin{figure*}
\center{\includegraphics[width=0.8\textwidth]{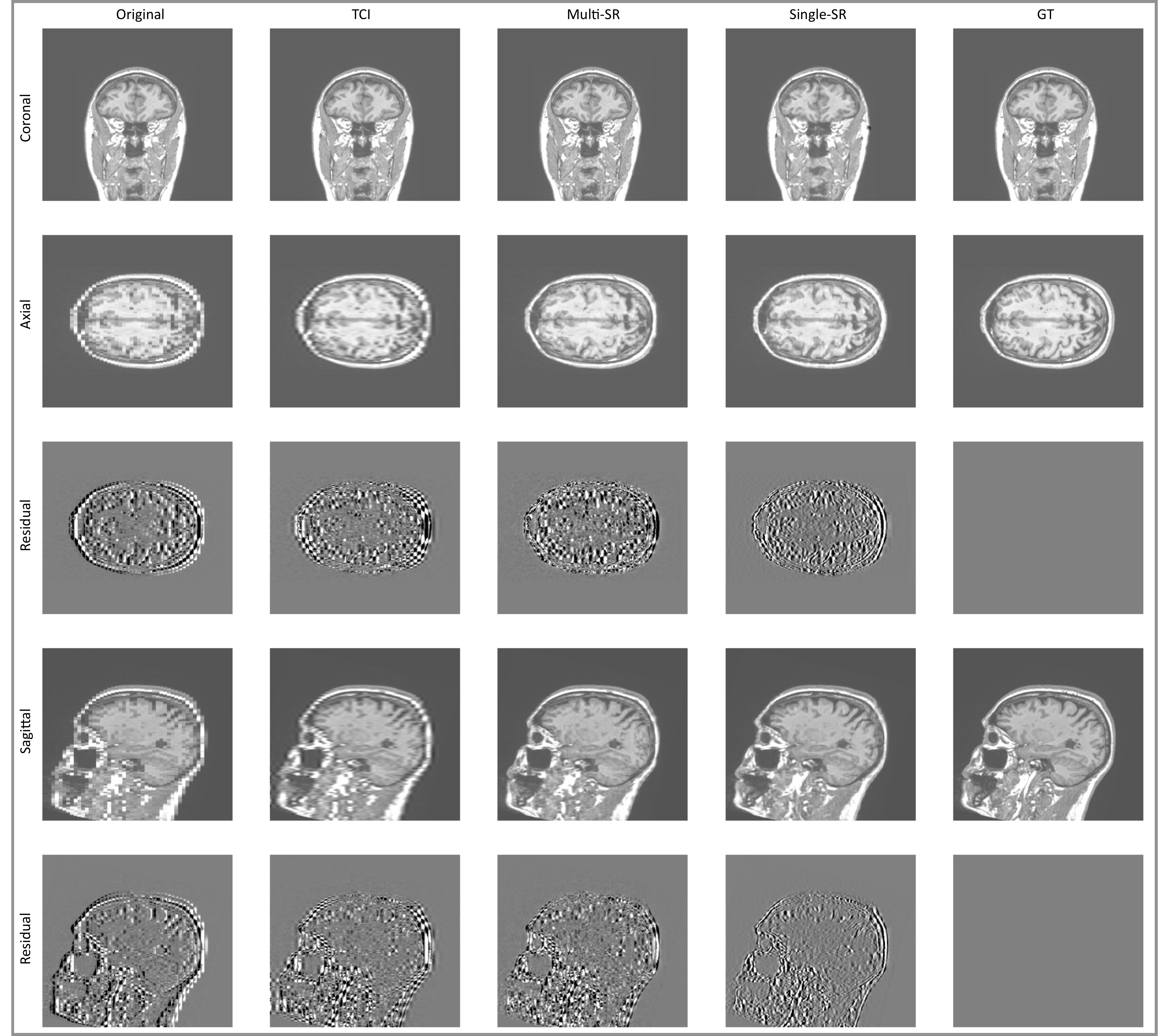}}
\caption{Comparison between single-scale model and multi-scale model on T1-weighted brain MRI. The figure also includes original slices, cubic interpolation and ground-truth slices as a reference. Residual images are added to better visualize the difference of the generated images to the ground-truths. 
}
\label{fig:4}
\end{figure*}

\subsection{GAN}    

In addition to the perceptual loss, we define a discriminator network $D_\theta$, together with the generator model $G_\theta$, based on the Generative Adversarial Networks structure. The generator $G_\theta$ and discriminator $D_\theta$ are optimized in an alternating manner to solve the adversarial min-max problem:
\begin{equation}
\min_G\max_D \mathbb{E}_{x\sim p_{\rm{gt}}}[\log D(x)] + \mathbb{E}_{y\sim p_{\rm{train}}}[\log(1-D(G(y)))],
\label{eq:5}
\end{equation}
which aims to train a generative model $G_\theta$ trying to fool the discriminator $D_\theta$, while the discriminator is trained simultaneously to distinguish the generated SR images from ground truth images. With this approach of adversarial training between $G_\theta$ and $D_\theta$, the generated images are encouraged residing in the featured manifold of real images, potentially with perceptually superior quality that is visually indistinguishable from the ground truth images.   

\section{Experiments}
\subsection{Training details}   

The training process is divided into two stages. The generator model $G_\theta$ is first trained by the MSE loss. The Adam optimization is used. The learning rate is initialized as $3 \times 10^{-4}$, and decayed by a factor of 3 upon three cycles, if the validation loss is no longer updated for the current cycle. Then, the model is further tuned by employing the perceptual loss and GAN. The total loss for $G_\theta$ upon the refined tuning is:
\begin{equation}
L_{G} = L_{per} + \lambda L_{GAN} + \mu L_{MSE},
\label{eq:6}
\end{equation}
where the weights $\lambda$ and $\mu$ are selected as 0.01 and 0.001. The pre-trained pixel-wise model is used to initialize the second stage of training. Transfer learning helps the perceptual-tuned GAN model converge more efficiently and obtain more visually pleasing results. The generator $G_\theta$ and the discriminator $D_\theta$ are alternately updated until the model converges to produce visually satisfactory images and no further improvement is visually apparent. 

\subsection{3D perceptual loss performance}

As discussed in section~\ref{sec:2}, pixel-wise loss usually achieves a relatively high metric score but fails in resolving the textures, edges, and high-frequency details, resulting in blurring of the images that are produced. The scheme that we propose generalizes the single-image perceptual loss based on pre-trained 2D VGG network into 3D, which is applicable for sliced medical datasets. 

As the first example of our qualitative results, Fig.~\ref{fig:3} shows the difference between using MSE loss and 3D perceptual loss with GAN, to train the network respectively. The two test datasets included in Fig.~\ref{fig:3} are the T1-weighted brain MRI (reformatted coronal images in the first row, and reformatted sagittal images in the second), and knee MRI in the third row. The images appear better compared to the original thick slices, the cubic-spline interpolated images, and the MSE results. They are further improved by 3D perceptual loss with GAN, achieving the highest perceptual quality.      

\subsection{Single-scale and multi-scale model comparison}

Fig.~\ref{fig:3} reveals the possibility of using SR technique as a better ``interpolation", given the active demand for data augmentation in medical-imaging deep-learning tasks. The feasibility of this idea requires an adaptive SR model, which is designed to incorporate random sampling factors. Fig.~\ref{fig:4} compares the single-scale SR model with our multi-scale SR model at the sampling factor equal to 5. We can imagine that, to some extent, the multi-scale model has to sacrifice partial resolutions, as a trade-off for allowing multiple scale entries. But still, the images obtained by the multi-scale model possess a satisfactory quality compared to the TCI and single-scale model. 

\begin{figure}
\center{\includegraphics[width=0.4\textwidth]{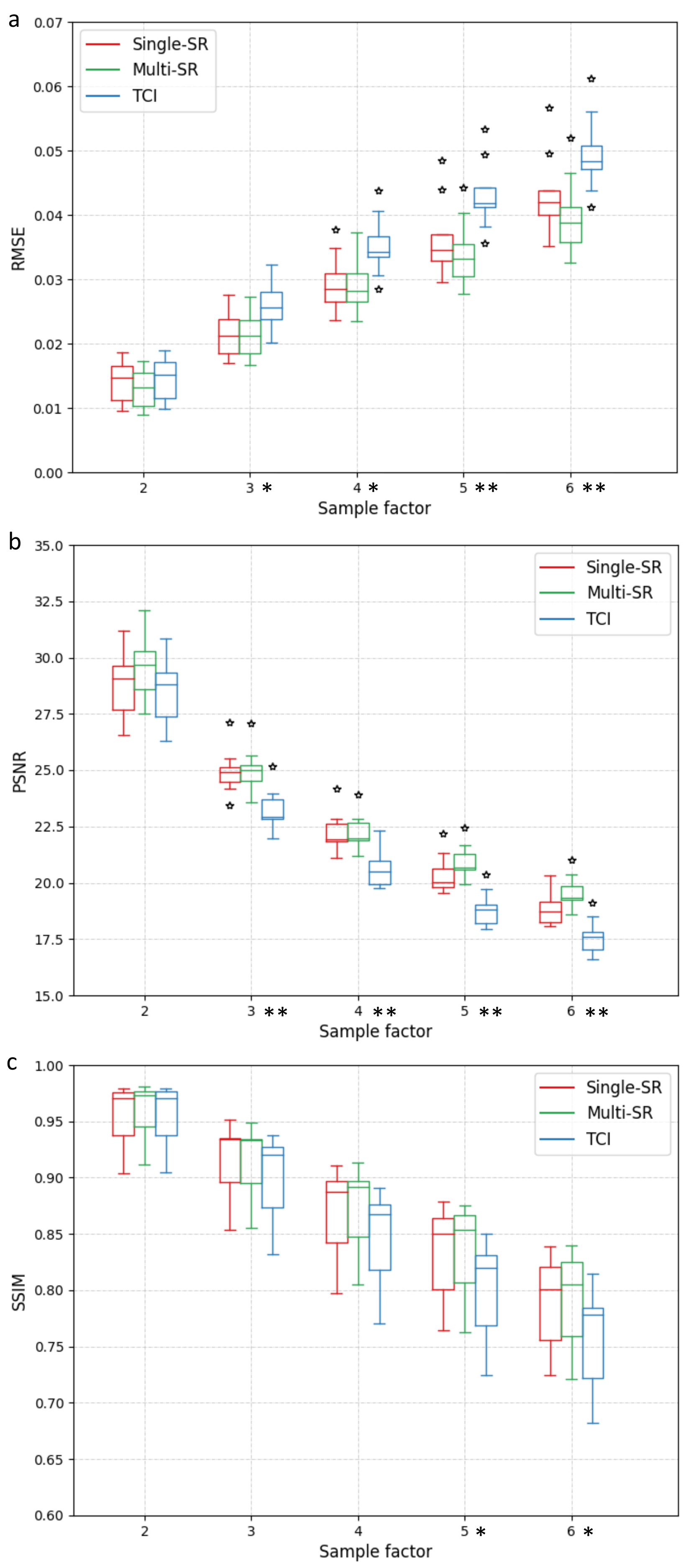}}
\caption{Evaluating on T1-weighted brain MRI test dataset with quantitative comparisons of measures: (a) RMSE, (b) PSNR and (c) SSIM in terms of the sampling factor s. Statistical significance is indicated on the x-axes with * for $p < 0.05$ and ** for $p < 0.001$.
}
\label{fig:5}
\end{figure}

\begin{figure*}
\center{\includegraphics[width=0.75\textwidth]{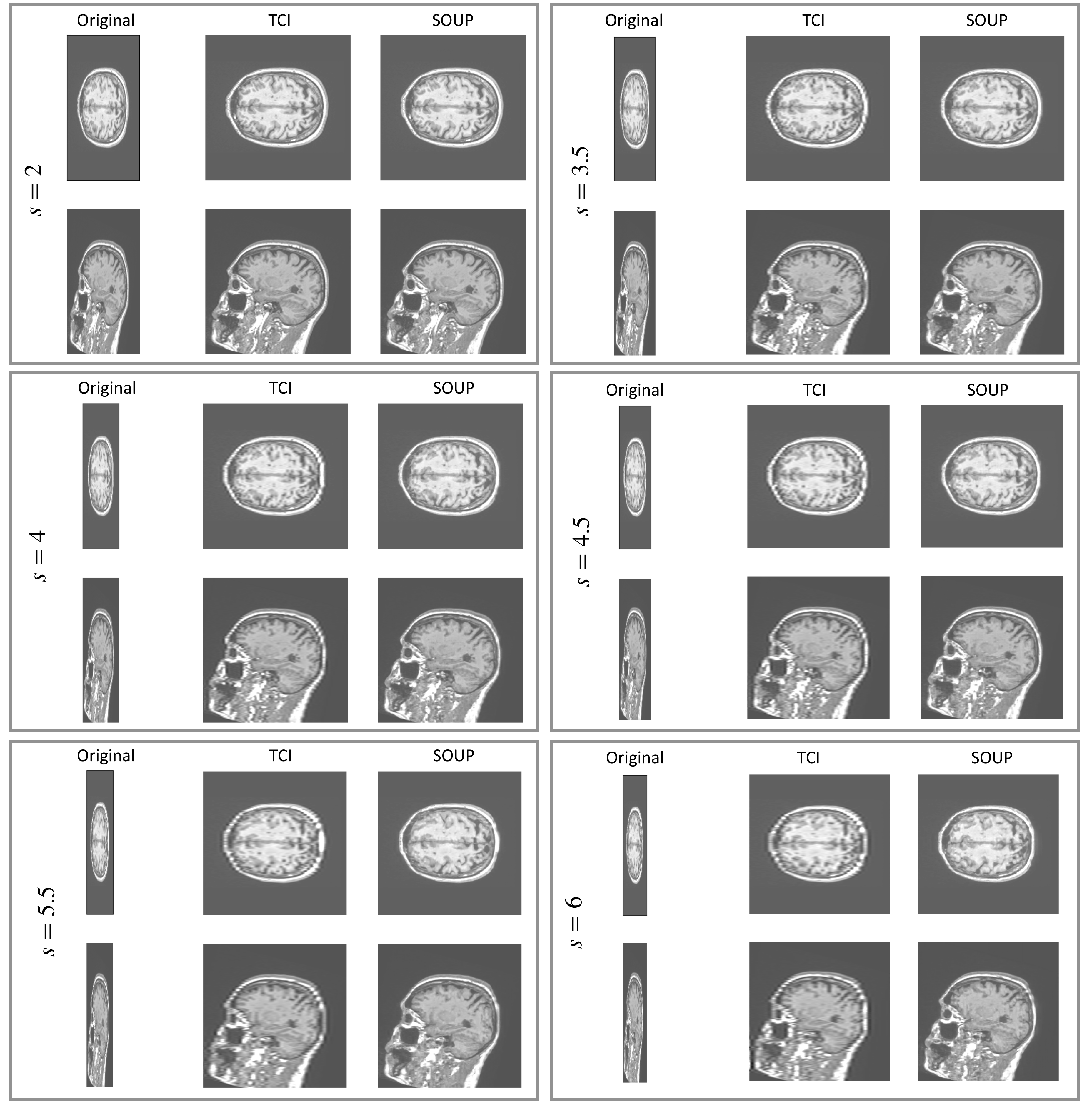}}
\caption{Comparison between multi-scale model and cubic interpolation at different scales. 
}
\label{fig:6}
\end{figure*}

\begin{figure*}
\center{\includegraphics[width=0.8\textwidth]{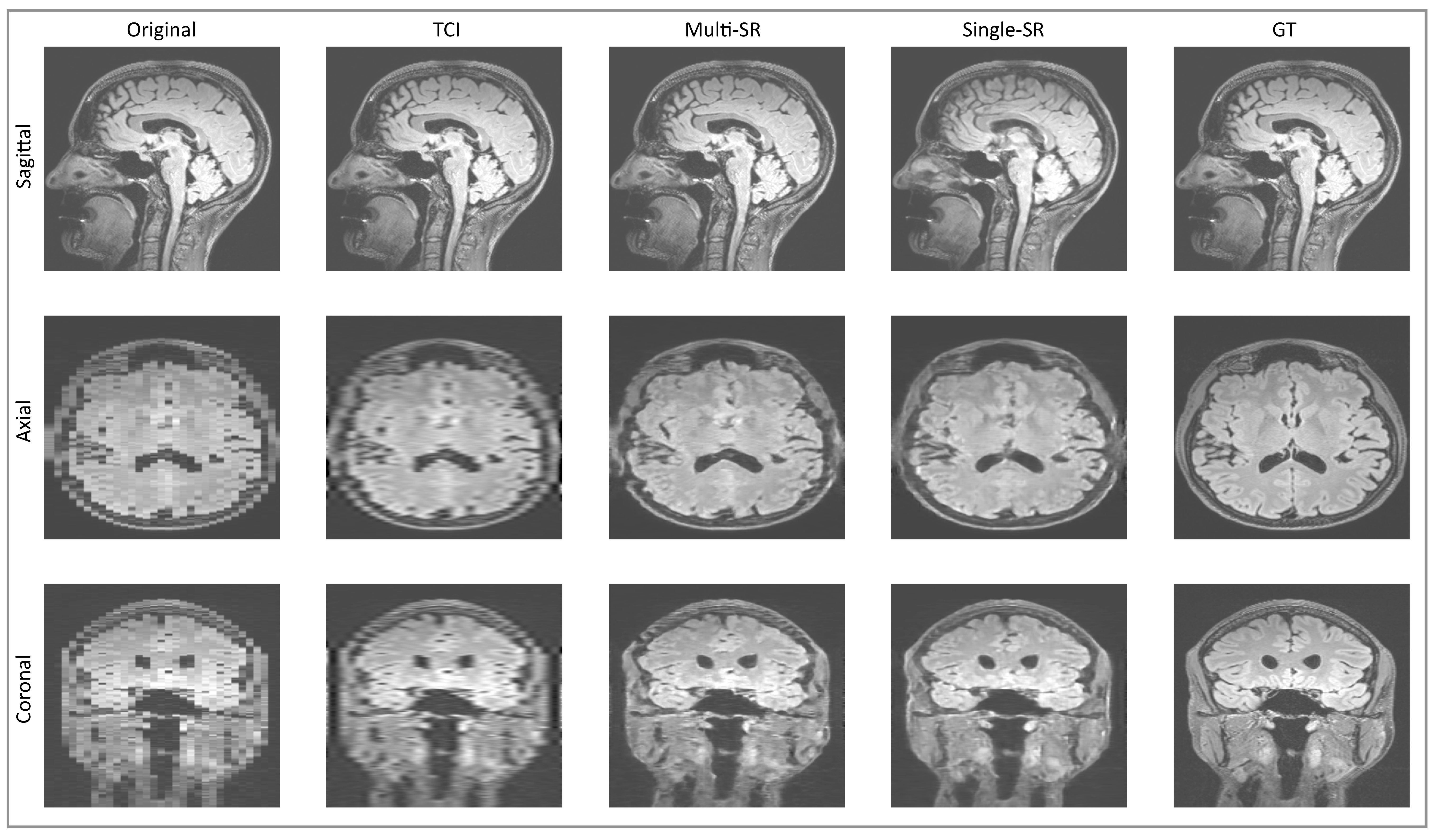}}
\caption{Comparison between single-scale model and multi-scale model performance on T2-FLAIR brain MRI.
}
\label{fig:7}
\end{figure*}

\begin{figure}
\center{\includegraphics[width=0.4\textwidth]{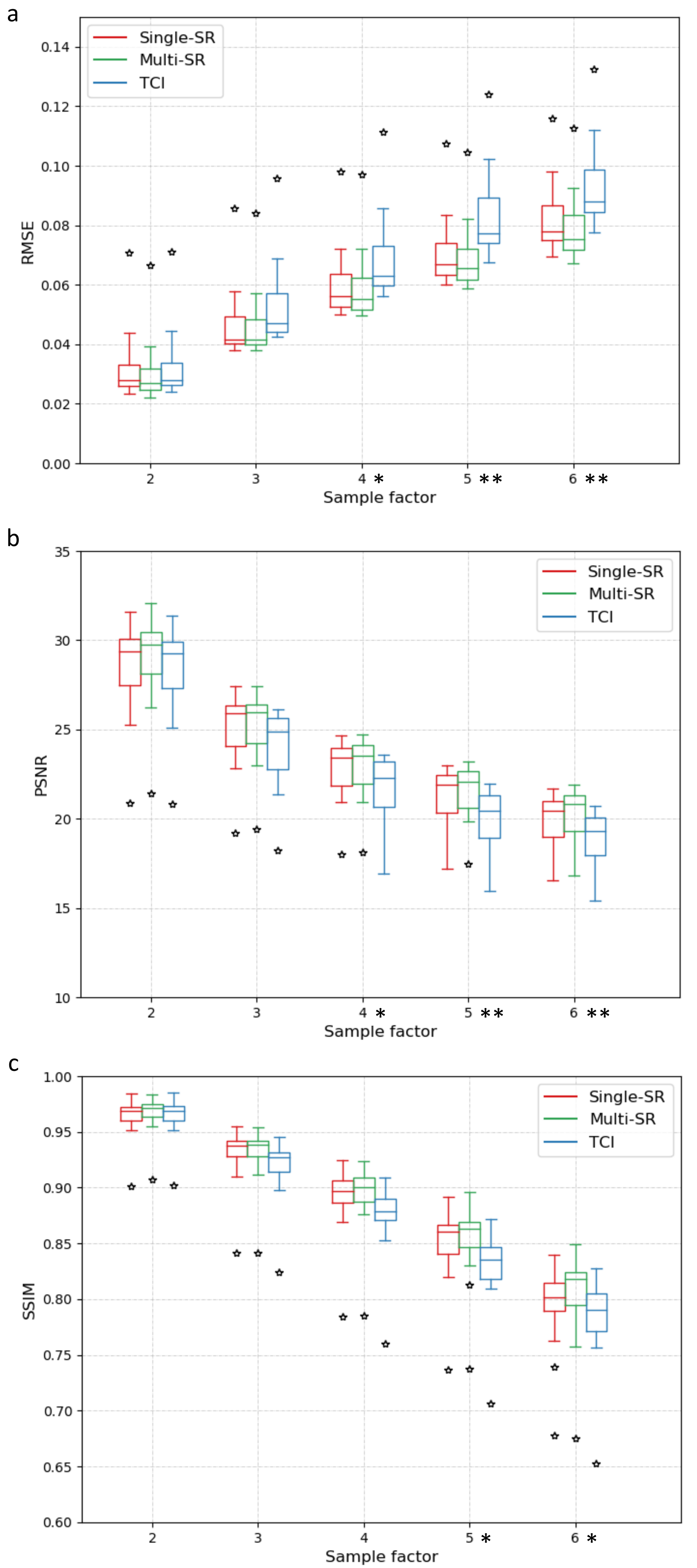}}
\caption{Evaluation using T2-weighted brain MRI images with quantitative comparisons of measures: (a) RMSE, (b) PSNR and (c) SSIM in terms of the sampling factor s. Statistical significance is indicated on the x-axes with * for $p < 0.05$ and ** for $p <0.001$.
}
\label{fig:8}
\end{figure}

\begin{figure}
\center{\includegraphics[width=0.5\textwidth]{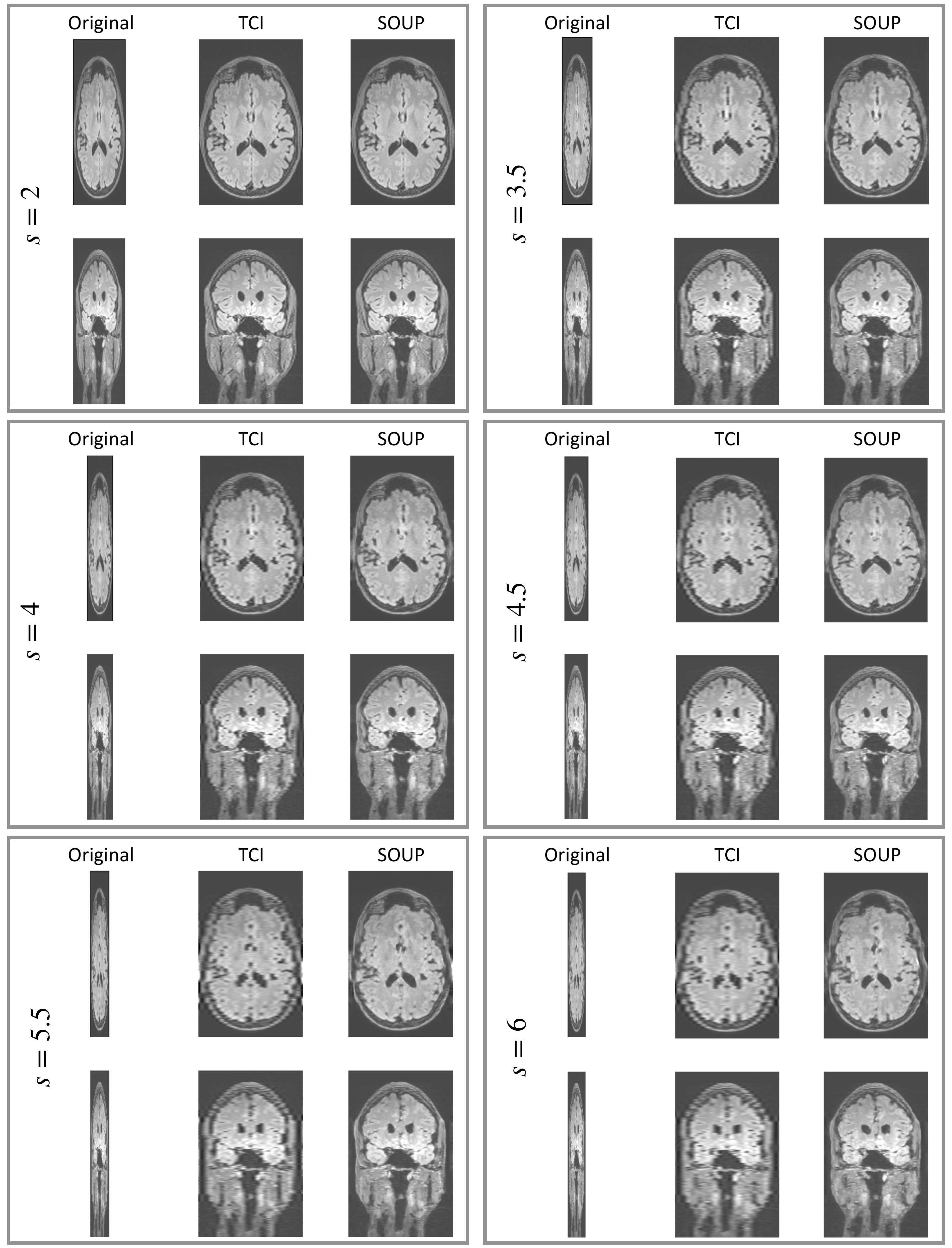}}
\caption{Comparison between multi-scale model and cubic interpolation on T2-FLAIR brain MRI at different scales. 
}
\label{fig:9}
\end{figure}
  
\begin{figure}
\center{\includegraphics[width=0.5\textwidth]{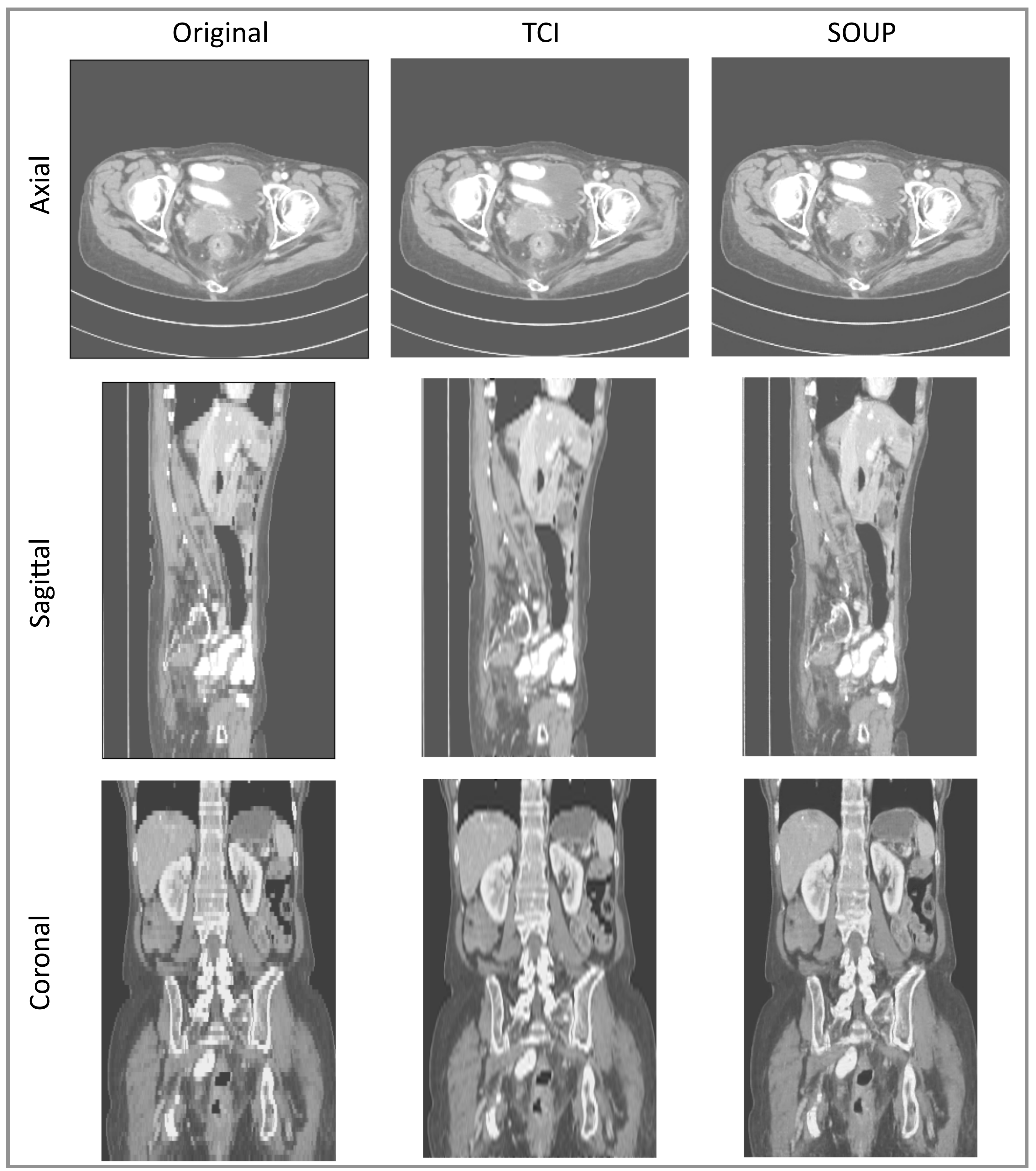}}
\caption{Example of SR interpolation on CT. Edges and textual details are better resolved by SR. Blurring is reduced.
}
\label{fig:10}
\end{figure}

Beyond the single-scale limit, our multi-scale model is demonstrated to work for user-selected scale values, which is the key to turning the SR model into a useful interpolation tool. Fig.~\ref{fig:5} reports the quantitative data performed on T1-weighted brain MRI test dataset with the metrics at different scales, including the root of mean squared error (RMSE), peak signal to noise ratio (PSNR)  and the structural similarity (SSIM). Fig.~\ref{fig:6} shows the generated images at different scales, obtained by SOUP (the multi-scale model) and compared to the conventional interpolation results. As a sum-up of the comparison, while the multi-scale SR model and the conventional interpolation method converge to a similar quality at the scale $s\le 2$, the SR model starts to outperform the interpolation and achieves better and better image qualities, when the scale $s$ increases.

\subsection{Application to other contrast types of MRI images}

In the literature, single image SR models are trained and tested on datasets of natural images, collected from different categories with various distributions. Similarly, an open question for medical images is whether it is possible to develop a generalized SR model that works for various medical image data types (e.g., modalities or contrast types) with an arbitrary sampling factor $s$ selected by the user. To address this question, we apply our multi-scale SR model to the test datasets of other medical imaging types, e.g., T2-FLAIR MRI and CT. Fig.~\ref{fig:7}  shows the comparison between single-scale model and multi-scale model on FLAIR brain MRI, Fig.~\ref{fig:8} quantitatively evaluates the performances of the models on the FLAIR MRI dataset with different metrics, and Fig.~\ref{fig:9} compares SOUP with conventionally interpolated images at different sampling factor $s$.

\begin{figure}
\center{\includegraphics[width=0.5\textwidth]{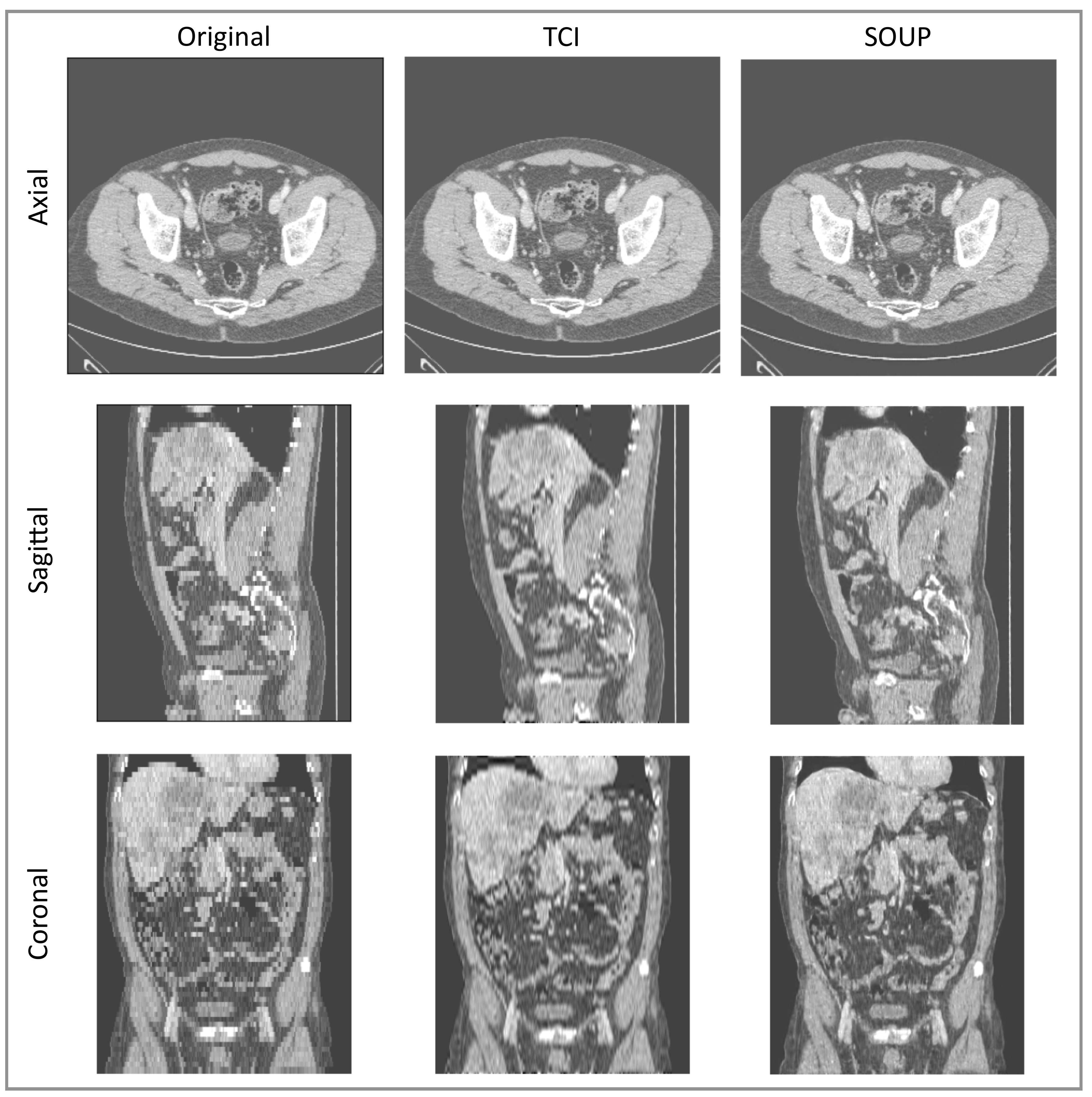}}
\caption{Another example of SR interpolation on CT data.
}
\label{fig:11}
\end{figure}

\subsection{Generalization to other medical imaging modalities, e.g., CT} 

We also apply our multi-scale SR model on CT data. Fig.~\ref{fig:10} and \ref{fig:11} show the SR-interpolated (SOUP) and conventionally interpolated images on abdominal and pelvic CTs. It is clear from both examples that the SR-interpolated results achieve significantly higher quality than the conventionally interpolated ones in regard to the edges, textual details and blurring. Those comparisons demonstrate our multi-scale SR model (SOUP) as a widely generalizable tool for different sampling factors and various applied imaging modalities, which paves the way for more advanced medical image interpolation through deep-learning SR.

\section{Conclusions}

To our knowledge, this is the first published work to combine the deep-learning SR method with a 3D perceptual-tuned GAN network, and to examine its application and generalization in terms of different sampling scales and various medical imaging modalities, as a novel 3D image interpolation technique. The superior-resolution thin slices obtained by the model outperform other conventional resolution-enhancement methods and previous SR work on medical images upon both qualitative and quantitative comparisons. Based on the promising simulation data, the proposed work shows potential in both clinical tasks, such as reducing the acquisition time and further resolving the MRI scans in retrospective studies, and research tasks, e.g., serving as a novel SR interpolation method for data augmentations with further applications in lesion measurements, radiomics and automatic segmentations.

%\section*{Acknowledgment}

%The preferred spelling of the word ``acknowledgment'' in American English is 
%without an ``e'' after the ``g.'' Use the singular heading even if you have 
%many acknowledgments. Avoid expressions such as ``One of us (S.B.A.) would 
%like to thank $\ldots$ .'' Instead, write ``F. A. Author thanks $\ldots$ .'' In most 
%cases, sponsor and financial support acknowledgments are placed in the 
%unnumbered footnote on the first page, not here.

\bibliographystyle{IEEEtran}
\bibliography{references}

\end{document}